# A Parametric Analysis of Project Management Performance to Enhance Software Development Process

Shashi Kumar N.R.
Advanced Software Engineering
Research Group, RIIC
Dayananda Sagar Institutions
Bangalore, India
nrshash@gmail.com

T.R. Gopalakrishnan Nair
Advanced Software Engineering
Research Group, RIIC, DSI
Aramco Endowed Chair -
Technology, PMU, KSA.
Bangalore, India
trgnair@ieee.org

Suma V
Advanced Software Engineering
Research Group, RIIC
Dayananda Sagar Institutions
Bangalore, India
sumavdsce@gmail.com

*Abstract*— Project Management process plays a significant role in effective development of software projects. Key challenges in the project management process are the estimation of time, cost, defect count, and subsequently selection of apt developers. Therefore precise estimation of above stated factors decides the success level of a project. This paper provides an empirical study of several projects developed in a service oriented software company in order to comprehend the project management process. The analysis throws light on the existence of variation in the aforementioned factors between estimation and observed results. It further captures the need for betterment of project management process in estimation and allocation of resources in the realization of high quality software product. The paper therefore aims to bring in an improved awareness in software engineering personnel concerning the magnitude and significance of better estimation and accurate allocation of resources for developing successful project.

*Keywords*- Software Process; Project Management Process; Software Engineering; Project Management; Software Quality; Defect Management

## I. INTRODUCTION

Software plays an essential role in the operation of wide spectrum of applications including banking to aircraft control systems. Sustainability of any software company depends on success rate of their projects which is achievable through development of customer satisfied quality products within estimated resources. Software by nature is however complex and hence development of high quality software becomes one of the challenging process. Since quality is not a state and is a continual process, it is important to engineer high quality artifacts during the development process [1].

A software process can be controlled, measured and improved by a project manager [2]. Authors in [3] have put forth the nature of volatility and how project managers can handle requirement volatility [3]. Further, defective design leads to failure of the application or at times become life intimidating [4].

Customer demands a reliable product within estimated schedule and budget. It is well known fact that software is highly people and their effort intensive which in turn modulates the success of a project. Project manager has an overall responsibility for the successful initiation, planning, design, execution, monitoring, controlling and closure of a project or a part of a project. Hence, the role of project manager becomes quite challenging to satisfy many factors such as customer needs, management goals, constraints within which the development team, maintenance team, testing team etc operates. Thus, the project manager plays a fundamental and significant role in success of a project and consequently success of the company. Hence, the process of developing and managing the project compels one to make realistic predictions and estimations to attain established target by optimal choice of available resources. However this research therefore focused on parametric analysis of project management performance to enhance software development process. The study comprises of deep investigations carried upon several empirical projects in a sampled software industry. The observations drawn from the study reflects variations in resource estimation and actual deployment which further affects the success of a project. The main objective of this paper is to analyze the level of accuracy in estimation and right allocation of resources towards realization of estimations during project development. Section II of the paper briefs about the background work for this investigation. Section III presents the research Methodology, Section IV provides the empirical analysis of several projects through a case study and Section V summarizes the paper.

## II. RELATED WORK

Author in [11] expresses that, specific knowledge, skills, efforts, experience, capabilities have to be combined in order to organize and manage a software development project successfully.

Author in [12] state that, in order to improve software organization's development processes and their capability to produce software products in a controllable and repeatable manner, various measurement programs have been developed.

The key challenge of process management is to continuously capture and apply the knowledge the software professionals possess [13].

Customer dissatisfaction, inadequate software quality, inability to deliver on time and within budget, and excessive rework are the issues required to be addressed to improve the software process.[14].

Effective software development is achievable through well-defined process and efficient developing team. Authors in [5] feel that the knowledge of project manager plays a vital role in the success or failure of projects [5].

Authors in [6] states that project managers and corporate management have critical role during software development process. They feel that software engineering has now taken a new perspective where project manager looks for professional management and software quality assurance methodologies during developmental activities [6].

However, author in [7] expresses that project manager has an upper hand in balancing and satisfying competing demands for project scope, time, cost, risk, and subsequently on quality [7].

Authors in [8] suggest that success of project depends much on project manager estimation capability of parameters like the number of defects and its presence at various phases of software development [8].

However, research made by authors in [9] indicates a vital need for analytical reasoning from project managers towards effective resource allocation for defect management in order to realize successful software projects [9].

## III. RESEARCH METHODOLOGY

This investigation includes a case study made in a service based software industry. Several projects were investigated to analyze the role and efficiency of project management process towards the realization of effective software development. In order to carry out the investigation, the Secondary data was collected from quality assurance department and from the document management center. Further, the mode of data collection includes log reports, face to face communication and interviews. The next steps of investigations lead towards analyzing the efficiency of project management process. Observation results indicate the need for enhancing the project management process in terms of accurate estimation and subsequent allocation of project influencing parameters and henceforth realization of development of a successful project.

## IV. CASE STUDY

The case study comprises of a CMMI level 5 and ISO certified service based software industry. The company functions on Business Intelligence, data warehouse, Enterprise Resource Planning, Business Process Outsourcing, Banking, Finance, Airlines and Energy Utilities.

Table 1 depicts a sampled data of twenty five projects developed since 2009 to 2012. The sampled projects are developed in .net and Java. The table provides information about the log of estimated and actual data of project influencing parameters such as project completion time, cost, total number of developers involved during developmental process and defect count.

The company follows several metrics for estimating time, cost and number of persons required for the development. The company estimates time based on two techniques namely complexity based estimation and Function Point Analysis [FPA].
In former case, time is estimated based on the categorization of tasks as simple, medium and complex tasks wherein task is primitive such as database creation. Time estimation for these tasks is based upon domain knowledge and experience of the project manager where a simpler task requires one person day which is eight constructive hours of a developer. Hence time required to complete the task therefore increases with increasing complexity.
In the latter method, irrespective of the technology, function points is calculated using number of inputs, outputs, interfaces, inquiries and complexity of the module [10]. The projects which are studied in this investigation therefore includes both the aforementioned techniques.

The metric used in the sampled projects for estimating cost which is also one of the project influencing parameter is:

$$Estimated\ Cost = No.of\ Estimated\ Total\ Project\ Hours \times 25\$ \qquad (1)$$

Where 25$ is the blended rate taken for estimation and if not the company uses the following rates are taken to estimate the cost:
Project Manager        -        30$/hour

Technical Lead       -   28$/hour
Senior Developer     -   26$/hour
Developer            -   23$/hour

The company further allocates project personnel whose effort is yet another modulating parameter by using the following the metric:

$$No.\ of\ persons\ estimated\ to\ complete\ project = \frac{Estimated\ Project\ in\ weeks}{40} \quad (2)$$

where 40 is a constant which is obtained as follows:

$$No.\ of\ working\ Hours\ per\ day \times No.\ of\ working\ days\ in\ a\ week = 8 \times 5 = 40 \quad (3)$$

Further defect is one of the most significant parameter which influences the quality of the project. The company in this case study estimates defect count based on the metric:

$$Total\ Estimated\ Defects = \frac{Total\ Estimated\ Time\ in\ Hours}{4.32} \quad (4)$$

TABLE 1: THE SAMPLED DATA OF 25 PROJECTS

| PROJECTS | TIME | | | COST($) | | | NO. OF DEVELOPERS | | | DEFECTS | | |
|---|---|---|---|---|---|---|---|---|---|---|---|---|
| | EST. | ACT. | % | EST. | ACT. | % | EST. | ACT. | % | EST. | ACT. | % |
| P1 | 131.84 | 128 | 2.9 | 3296 | 3200 | 2.9 | 1 | 1 | 0.0 | 36 | 30 | 15.6 |
| P2 | 217.5 | 174 | 20.0 | 5437.5 | 4350 | 20.0 | 1 | 2 | 100.0 | 48 | 45 | 6.9 |
| P3 | 278.4 | 232 | 16.7 | 6960 | 5800 | 16.7 | 1 | 1 | 0.0 | 64 | 67 | -4.0 |
| P4 | 288 | 240 | 16.7 | 7200 | 6000 | 16.7 | 1 | 1 | 0.0 | 67 | 65 | 2.5 |
| P5 | 297.6 | 248 | 16.7 | 7440 | 6200 | 16.7 | 1 | 1 | 0.0 | 69 | 72 | -4.5 |
| P6 | 432 | 360 | 16.7 | 10800 | 9000 | 16.7 | 1 | 2 | 100.0 | 100 | 95 | 5.0 |
| P7 | 582.4 | 520 | 10.7 | 14560 | 13000 | 10.7 | 1 | 1 | 0.0 | 144 | 145 | -0.4 |
| P8 | 700.8 | 584 | 16.7 | 17520 | 14600 | 16.7 | 1 | 2 | 100.0 | 162 | 159 | 2.0 |
| P9 | 907.2 | 720 | 20.6 | 22680 | 18000 | 20.6 | 1 | 2 | 100.0 | 200 | 201 | -0.5 |
| P10 | 792 | 720 | 9.1 | 19800 | 18000 | 9.1 | 1 | 2 | 100.0 | 200 | 198 | 1.0 |
| P11 | 836 | 760 | 9.1 | 20900 | 19000 | 9.1 | 1 | 2 | 100.0 | 211 | 215 | -1.8 |
| P12 | 912 | 760 | 16.7 | 22800 | 19000 | 16.7 | 1 | 2 | 100.0 | 211 | 215 | -1.8 |
| P13 | 1472 | 1280 | 13.0 | 36800 | 32000 | 13.0 | 1 | 2 | 100.0 | 356 | 349 | 1.8 |
| P14 | 1478.4 | 1320 | 10.7 | 36960 | 33000 | 10.7 | 1 | 2 | 100.0 | 367 | 353 | 3.7 |
| P15 | 2017.2 | 1640 | 18.7 | 50430 | 41000 | 18.7 | 2 | 3 | 50.0 | 456 | 459 | -0.8 |
| P16 | 2132.8 | 1720 | 19.4 | 53320 | 43000 | 19.4 | 2 | 2 | 0.0 | 478 | 481 | -0.7 |
| P17 | 2313.6 | 1928 | 16.7 | 57840 | 48200 | 16.7 | 2 | 3 | 50.0 | 536 | 535 | 0.1 |
| P18 | 2332 | 2120 | 9.1 | 58300 | 53000 | 9.1 | 2 | 2 | 0.0 | 589 | 595 | -1.0 |
| P19 | 4404.4 | 3640 | 17.4 | 110110 | 91000 | 17.4 | 3 | 5 | 66.7 | 1011 | 1100 | -8.8 |
| P20 | 4200 | 4000 | 4.8 | 105000 | 100000 | 4.8 | 3 | 5 | 66.7 | 1111 | 1200 | -8.0 |
| P21 | 4569.6 | 4080 | 10.7 | 114240 | 102000 | 10.7 | 3 | 3 | 0.0 | 1133 | 1120 | 1.2 |
| P22 | 7344 | 6120 | 16.7 | 183600 | 153000 | 16.7 | 5 | 7 | 40.0 | 1700 | 1665 | 2.1 |
| P23 | 7599.2 | 6440 | 15.3 | 189980 | 161000 | 15.3 | 5 | 7 | 40.0 | 1789 | 1652 | 7.7 |
| P24 | 7700 | 7000 | 9.1 | 192500 | 175000 | 9.1 | 6 | 8 | 33.3 | 1944 | 1532 | 21.2 |
| P25 | 8592 | 7160 | 16.7 | 214800 | 179000 | 16.7 | 6 | 7 | 16.7 | 1989 | 1851 | 6.9 |

EST.- Estimated; ACT.-Actual

Where 4.32 is a constant obtained by multiplying the constants such as

$$4.32 = 8 \times 1.8 \times 0.3 \quad (5)$$

Where 8 is the No. of hours in a day
1.8 is the constant for .Net and java Projects
0.3 is a constant

The constant 4.32 is obtained by following the computations as given below:

- Initially, number of hours is divided by 8 constructive working hours to obtain total number of days which is further divided by a constant 1.8.
- The constant is evaluated by the company from their historical study for java/ .net projects which results in estimation of total number of function points
- Defects are estimated based upon the function point calculations
- Therefore 4.32 is arrived using number of function points which depicts the complexity of a software

The Table 1 infers that estimation of total developmental time and actual development time varies up to 20%. Estimation of cost and actual expense incurred for the resources throughout the development of the project also varies up to 20%. Huge variations of almost nearing to double the need of estimated number of developing personnel are observed in several projects and are also depicted in the sampled Table 1. However, defect count estimation and the captured defect count are observed to vary in a highly unstable way.

The above inferences are further depicted in the form of graph which is plotted for each parameter which influences the software development process significantly.

Figure 1 illustrates the estimated project development time and actual project developmental time measured in person hours. Figure 2 illustrates the estimated total project cost and actual project budget incurred. Figure 3 depicts the estimated number of resource personnel for the project development and actual developers involved during the complete project developmental activities. Figure 4 illustrates estimated defect count in the entire project and actual defect count identified before the product release. Figure 5 depicts the variation between estimation and actual observations of the aforementioned factors in terms of percentage.

The observational inferences indicates that despite stringent implementation of metrics for estimation purpose, the allocation of right choice of resources is highly immature and results in compromised software product in lieu of customer anticipated software product. This further indicates a need for enhancing the project management process through analytical mode rather than intuitive mode which are conventional till now. Our forthcoming research explores rationale for the observed behavior in the development process from the perspective of project management process and impact analysis on the same.

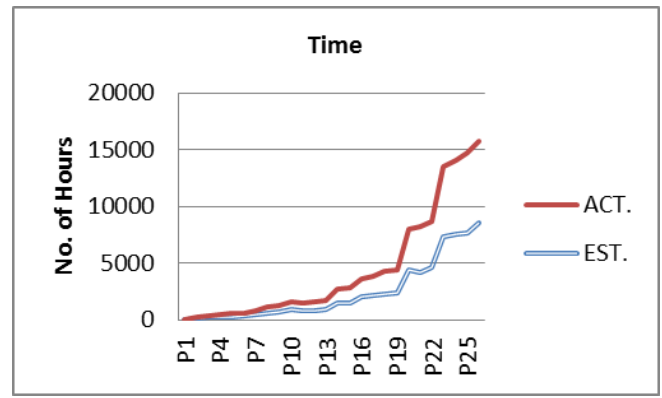

Figure 1. Estimated project development time and actual project developmental time in hours

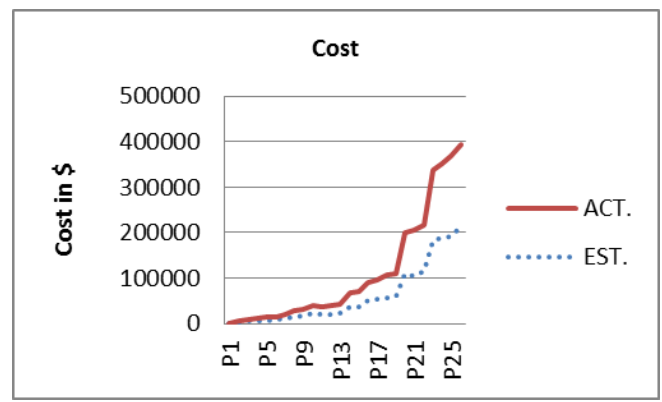

Figure 2. The estimated total project cost and actual project budget incurred.

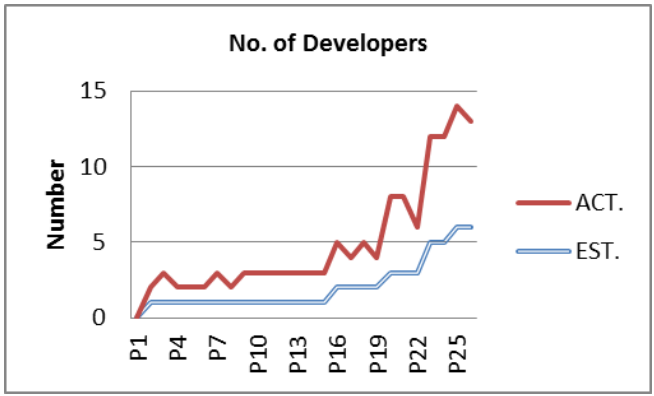

Figure 3. The estimated number of resource personnel for the project development and actual developers involved

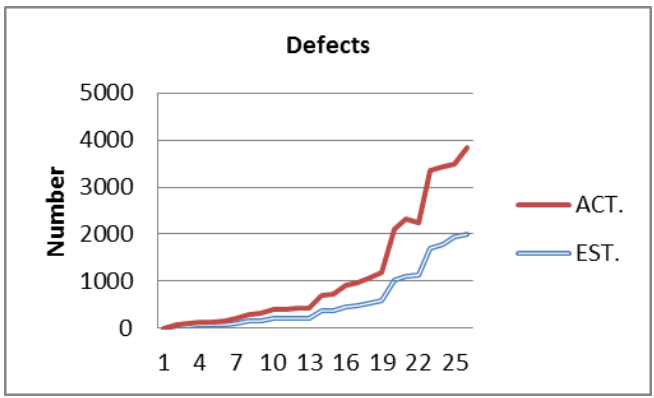

Figure 4. Estimated defect count in the entire project and actual defect count

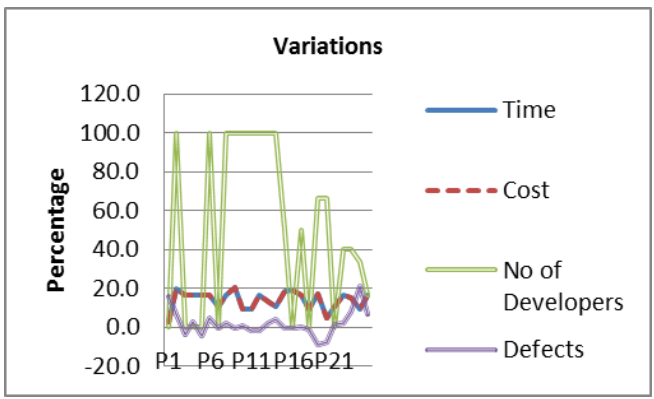

Figure 5. The variation between estimation and actual observations

## V. CONCLUSION

Development of a customer satisfied software products is the core requirement of any software industry which is achievable through the production of high quality software. Quality is realized by developing the product within the estimated resource constraints.

This paper presented a case study of a leading software industry where several empirical projects were investigated in order to analyze the efficiency of project management process in terms of significant project influencing parameters such as time, cost, no. of developers and defect count.

The observational results indicate the need for the enhancement of existing software project management process through analytical mode.

This awareness further enables one to improve the quality of the project and in turn increased customer satisfaction. Additionally it improves the productivity and organizational business growth.

ACKNOWLEDGMENT

The authors would like to acknowledge the software company involved in this study and the project managers for their invaluable help in providing necessary information for our work under the framework of the Non-Disclosure Agreement.